\documentclass[prl,twocolumn,showpacs,amsmath,amssymb,superscriptaddress]{revtex4}

\usepackage{graphicx}
\usepackage{dcolumn}
\usepackage{bm}
\usepackage{bbm}
\usepackage{color}

\begin{document}

\title{Prediction of Near-Room-Temperature Quantum Anomalous Hall Effect  on Honeycomb Materials
}
\author{Shu-Chun Wu}
\affiliation{Max Planck Institute for Chemical Physics of Solids, D-01187 Dresden, Germany}
\author{Guangcun Shan}
\affiliation{Max Planck Institute for Chemical Physics of Solids, D-01187 Dresden, Germany}
\author{Binghai Yan}\email{yan@cpfs.mpg.de}
\affiliation{Max Planck Institute for Chemical Physics of Solids, D-01187 Dresden, Germany}
\affiliation{Max Planck Institute for Physics of Complex Systems, D-01187 Dresden, Germany}

\begin{abstract}
Recently, this long-sought quantum anomalous Hall effect was realized in the magnetic topological insulator. However, the requirement of an extremely low temperature (approximately 30~mK) hinders realistic applications. Based on \textit{ab-initio} band structure calculations, we propose a quantum anomalous Hall platform with a large energy gap of 0.34 and 0.06 ~eV on honeycomb lattices comprised of Sn and Ge, respectively. The ferromagnetic order forms in one sublattice of the honeycomb structure by controlling the surface functionalization rather than dilute magnetic doping, 
{\color{black}which is expected to be visualized by spin polarized STM in experiment}. Strong coupling between the inherent quantum spin Hall state and ferromagnetism results in considerable exchange splitting and consequently an ferromagnetic insulator with a  large energy gap. The estimated mean-field Curie temperature 
is 243 and 509~K for Sn and Ge lattices, respectively. The large energy gap and high Curie temperature indicate the feasibility of the quantum anomalous Hall effect in the near-room-temperature and even room-temperature regions. 
\end{abstract}
\maketitle

The quantum anomalous Hall (QAH) effect is a topologically nontrivial phase characterized by a finite Chern number and chiral edge states inside the bulk band gap, which leads to the quantized Hall effect without a magnetic field~\cite{Haldane1988}. The chiral edge states carry dissipationless electric current owing to robustness against backscattering ~\cite{Halperin1982} and are therefore attractive for applications in low-power-consumption electronics. Recently, this long-sought QAH effect was realized chromium-doped (Bi,Sb)$_2$Te$_3$ (ref.~\cite{Chang2013}), the magnetic topological insulator~\cite{qi2011RMP,hasan2010}. However, the requirement of an extremely low temperature (approximately 30~mK) hinders realistic applications, which is fundamentally limited by the bulk energy-gap and the ferromagnetic Curie temperature~\cite{Chang2013}. Thus, novel materials are in high demand for future applications of the QAH effect.

In a quantum spin Hall (QSH) system~\cite{kane2005B,bernevig2006d,koenig2007}, the topological band structure is identified by a band inversion in both spin channels that are time-reversal (TR) conjugates of each other. A pair of counter-propagating edge states with opposite spins exists, i.e., two copies of quantum anomalous Hall (QAH) edge states. If the ferromagnetic (FM) order suppresses one of the spin channels, it can lead to the QAH effect~\cite{qi2006b,liu2008,yu2010}. The band inversion in a single spin channel, which is characterized by a finite Chern number, originates from the gapless edge state inside the bulk energy-gap. Therefore, the following transition-metal-doped topological insulators were proposed to realize the QAH effects: the aforementioned chromium-doped Bi$_2$Te$_3$ (ref.\cite{yu2010}), manganese-doped HgTe quantum wells (QWs)~\cite{liu2008,Hsu2013}, and other magnetically doped QWs~\cite{Wang2013,Zhang2014}. The bulk energy-gap of these 2D QSH insulators are usually of the order of 1 or 10 meV.
It is believed that the ferromagnetism of a magnetic QSH insulator can be enhanced owing to the band edge singularity of such an inverted band structure when the Fermi energy ($E_{\rm F}$) lies inside the gap~\cite{yu2010,Wang2013}, despite the fact that free carriers are required to mediate the coupling between magnetic moments to form the FM order in a common dilute magnetic semiconductor. For example, chromium-doped Bi$_2$Te$_3$ exhibits an FM order with Curie temperature $T_{\rm C} \approx 15 $ K, while manganese-dope HgTe is paramagnetic~\cite{Nagata1980} in experiment.
In addition, QAH states have also been predicted to exist in transition metal oxide heterostructures~\cite{Xiao2011,Ruegg2011,Yang2011,Wang2011,Hu2012,Ruegg2012}, graphene~\cite{Qiao2010,Tse2011,Qiao2014}, silicene~\cite{Liu2011silicence,Liu2011prb,Ezawa2012}, magnetic thin layers containing heavy elements~\cite{Garrity2013}, and magnetic topological crystalline insulators~\cite{Fang2014,Zhang2013}.

In this letter, we propose the realization of the QAH effect on graphene-like honeycomb lattices comprised of Sn and Ge, called stanene and germanene, respectively. Recently, they were found to be large energy gap QSH insulators by band structure calculations~\cite{Xu2013,Si2014} and model . We introduce ferromagnetism on these 2D lattices by controlling the surface functionality instead of transition metal doping, wherein one sublattice is fully passivated by halide atoms while the other is not. Based on \textit{ab-initio} calculations, the FM order drives the QSH phase to QAH phase that exhibits a large energy gap of 0.34 and 0.06 ~eV for stanene and germanene, respectively. We also discuss the possible experimental realization on a semiconductor substrate.

The \textit{ab-initio} density-functional theory (DFT) calculations have been performed using the projector augmented wave method, which was implemented in the Vienna \textit{ab initio} simulation package (\textsc{vasp}) \cite{kresse1996}. Hybrid-functional method (HSE06)~\cite{Heyd2003,Heyd2006} was adopted in the electronic structure total energy calculations to avoid {\color{black}the self-interaction problem} of the local density approximations (LDA), which was essential to reveal correct electronic properties in current systems.
All atomic positions and lattice parameters are fully optimized before the electronic structure calculations. The DFT Bloch wave functions were further projected to maximally localized Wannier functions \cite{Marzari1997}. Based on Wannier functions, the anomalous Hall conductivity $\sigma_{xy}$ were calculated using the Kubo-Greenwood formula. The CdTe substrate was simulated by one CdTe atomic layer, wherein the bottom dangling bonds of Cd atoms are passivated by Cl. Edge states, i.e. local density of states that was projected to the edge of a semi-infinite plane, were calculated using the iterative Green function method. The edge configuration is a zigzag-type boundary of a honeycomb lattice. {\color{black} Spin-orbit coupling (SOC) were included in charge self-consistent DFT calculations. The magnetic ground states were deduced by comparing the total energies between different magnetic configurations.}

\begin{figure}
\centering
\includegraphics[width = 1\columnwidth]{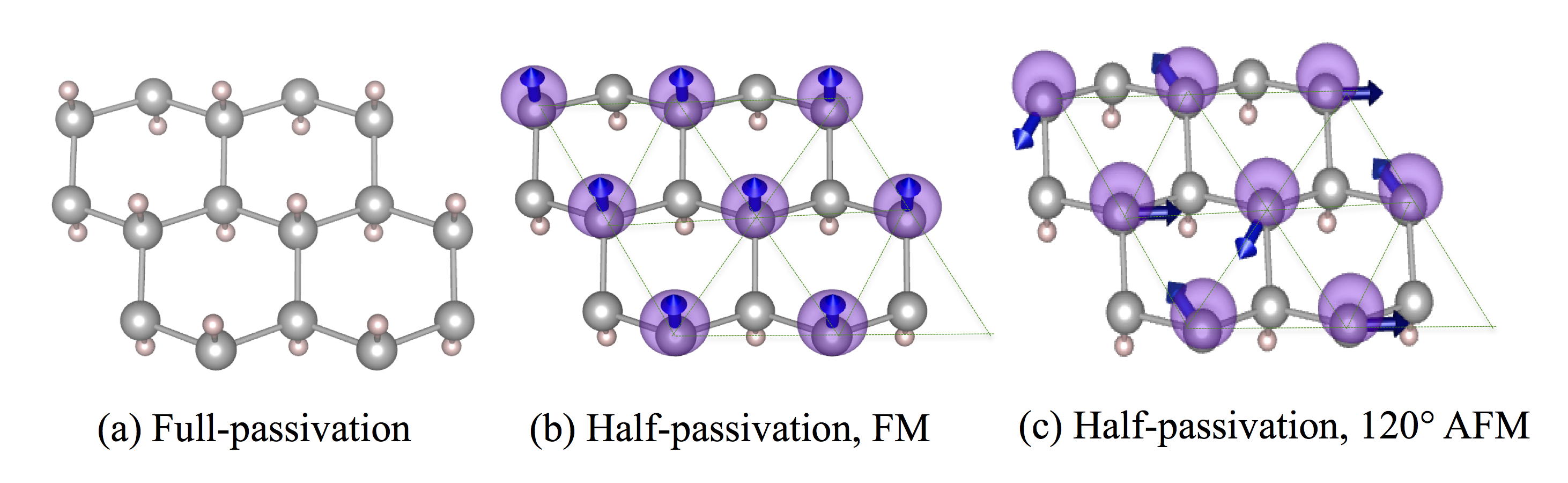}
  \caption{(color online) Structures of stanene with (a) full- and (b)-(c) half-passivation by I atoms. Large gray and small pink spheres represent the Sn and I atoms, respectively. For half-passivated stanene, unpassivated Sn sites exhibit magnetic moments in a triangular lattice (dotted lines), which are indicated by blue arrows. The FM and 
 120$^{\circ}$  AFM phases are indicated in (b) and (c), respectively.
 The spin charge density is shown in an isovalue surface plot, which lies primarily at the unpassivated Sn atoms. 
 The isovalue of the spin density is chosen as 0.006~electron/\AA$^3$.}
\label{fig:structure}
\end{figure}

\begin{figure}
\centering
\includegraphics[width = 1\columnwidth]{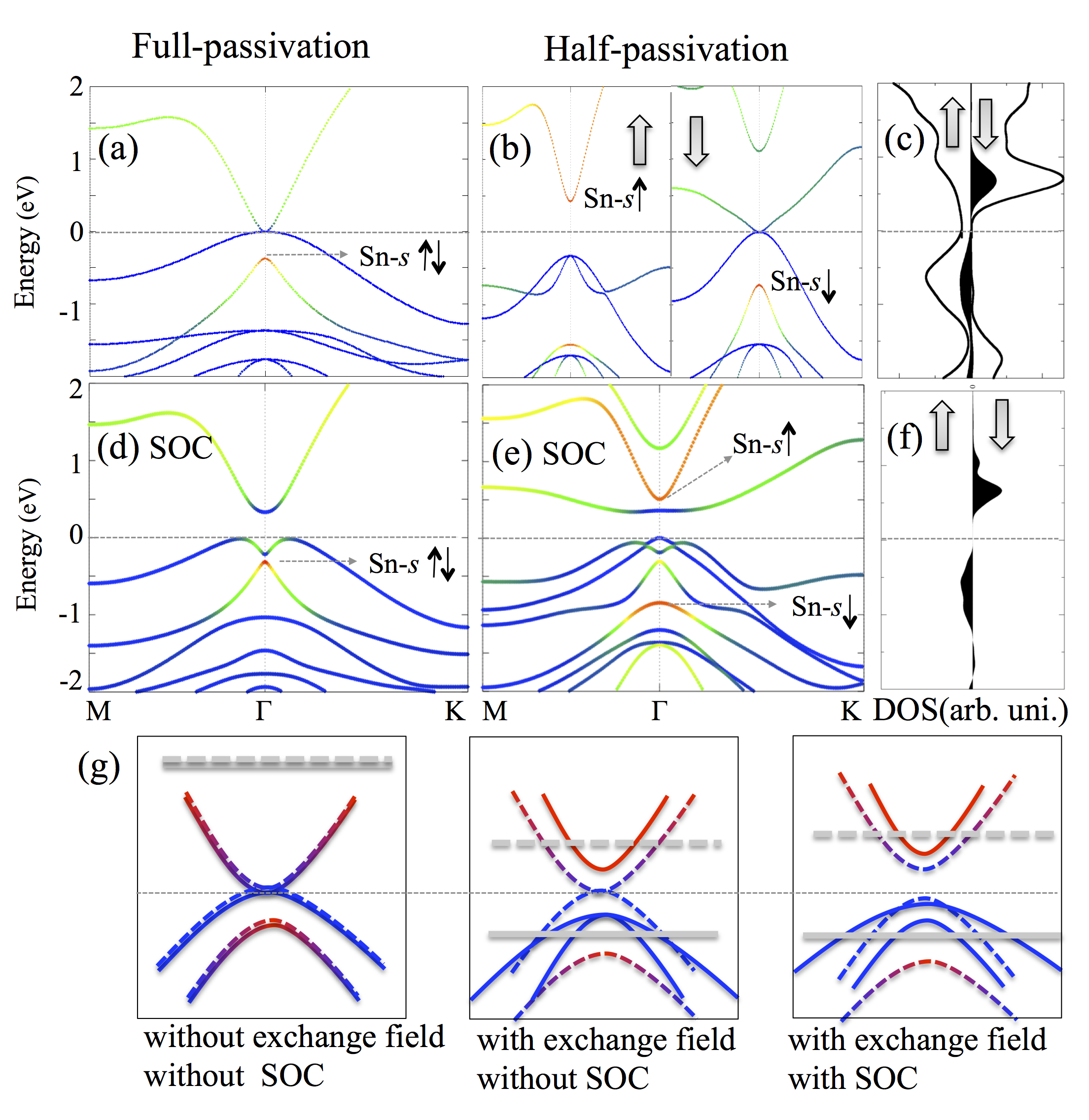}
  \caption{(color online)  Bulk band structures. The upper (a)-(c) and middle (d)-(f) panels show band structures calculated without and with SOC. 
  The left panel (a) \& (d) shows the fully passivated case, while the middle and right panels (b),(c),(e) \& (f) show the half-passivated case. 
  In band structures, color from blue to red represents the increasing component of Sn-$s$ orbitals, which is used to indicate the band inversion. 
  The Fermi energy is shifted to zero, as indicated by horizontal gray lines. 
  (c) and (f) are DOS correspond to (b) and (e), respectively, 
  in which the empty black curves represent total DOS and filled black curve represent the projection to the Sn-$p_z$ dangling bond state.
  (g) Illustration of the band evolution from the QSH to QAH states.
  The red and blue lines represent $s$- band $p$-bands, respectively. The $p_z$ states are illustrated by gray horizontal lines. 
  The thick solid and dashed lines represent different spin channels.
  }
\label{fig:band}
\end{figure}

Stanene and germanene are graphene lattices that are functionalized by halide atoms. Herein, we use stanene as an example. The in-plane $\sigma$-bonds connect the honeycomb lattice where two Sn sublattices are slightly buckled oppositely out-of-plane. The band structure of the Sn pristine honeycomb lattice is equivalent to that of graphene, in which unpaired Sn-$p_z$ orbitals form Dirac-type states at the K (K$'$) points~\cite{Xu2013}. When halide atoms (Cl, Br, or I) are adopted to passivate Sn-$p_z$ states (see Fig.~\ref{fig:structure}a), the Dirac-type bands are eliminated and Sn-$sp$ states emerge at the $\Gamma$ point as the low-energy bands near $E_{\rm F}$. The QSH phase appears owing to an inversion between the Sn-$s$ and Sn-$p_{xy}$ bands, as shown in Fig.~\ref{fig:band}d. {\color{black} We note that the $s$-band is an anti-bonding state between two Sn atoms and thus exhibit ``-'' in parity, while the $p_{xy}$-band is a bonding state and thus shows ``+'' parity ~\cite{SM}. The lowest conduction band and the highest valence band are the SOC split from $p_{xy}$-bands, $j=3/2$ and $j=1/2$ states, respectively. Here the $s-p_{xy}$ inversion refers to the inverted order between the $s$-band and the $j=3/2$ band.}
If the halide atom is removed from a single Sn site, the resultant unpassivated Sn-$p_z$ electron acts as a localized spin-1/2 site. When a complete Sn sublattice exhibits dangling bonds while the other sublattice is still passivated, e.g. by removing all the halide atoms on the top surface, magnetic moments due to unpaired Sn-$p_z$ electrons form a triangular lattice and possibly spontaneously evolve into an FM order, as shown in Fig.~\ref{fig:structure}b.


 \begin{figure}
\centering
\includegraphics[width = 1\columnwidth]{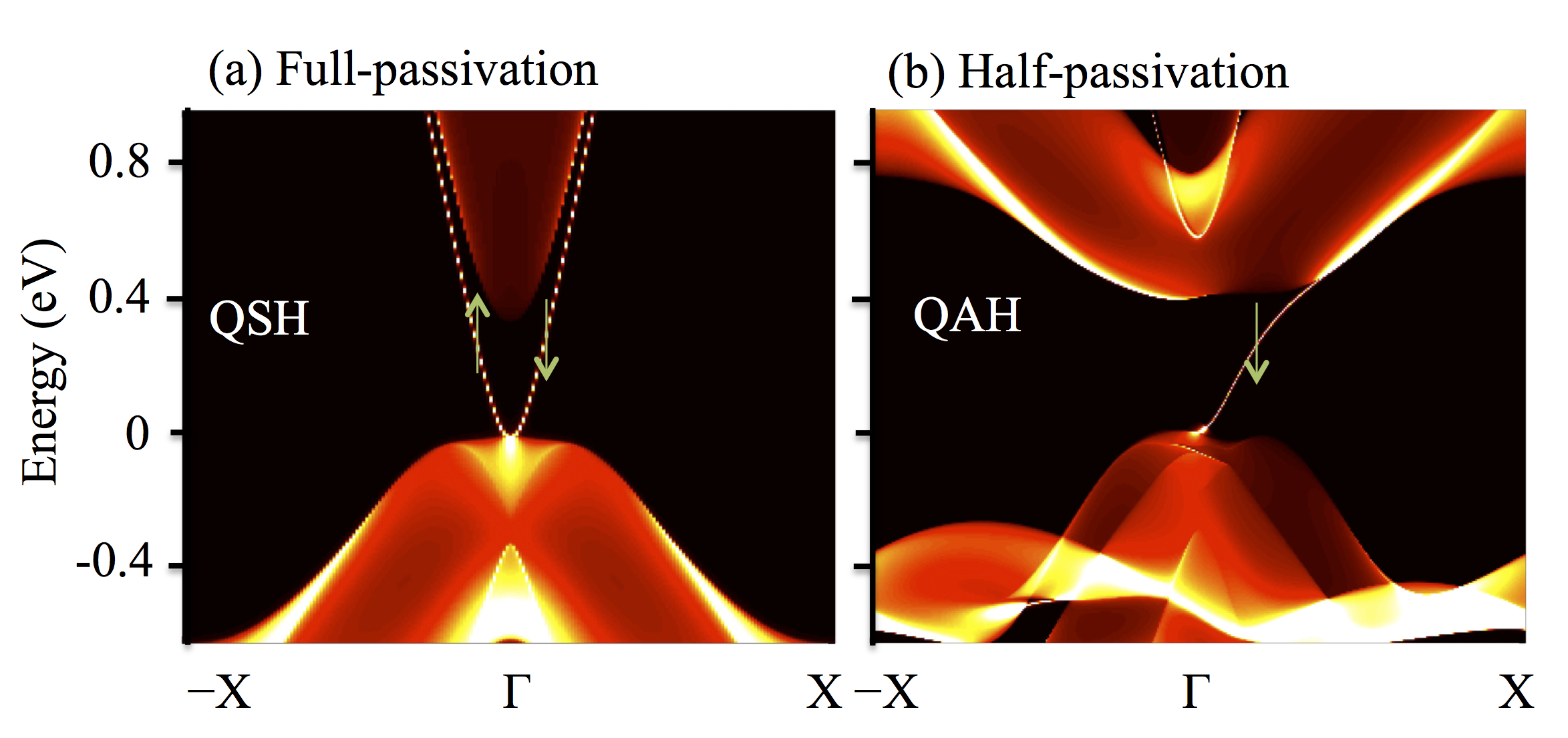}
  \caption{(color online) Calculated local density of states of edge states for (a) QSH and (b) QAH insulators.  
  The edge states are calculated on the edge of a semi-infinite plane. The warmer colors (white) represent higher local density of states, while the red (light gray) and black regions indicate 2D bulk energy bands and energy gaps, respectively.}
  \label{fig:edge}
\end{figure}

\begin{figure}
\centering
\includegraphics[width = 1\columnwidth]{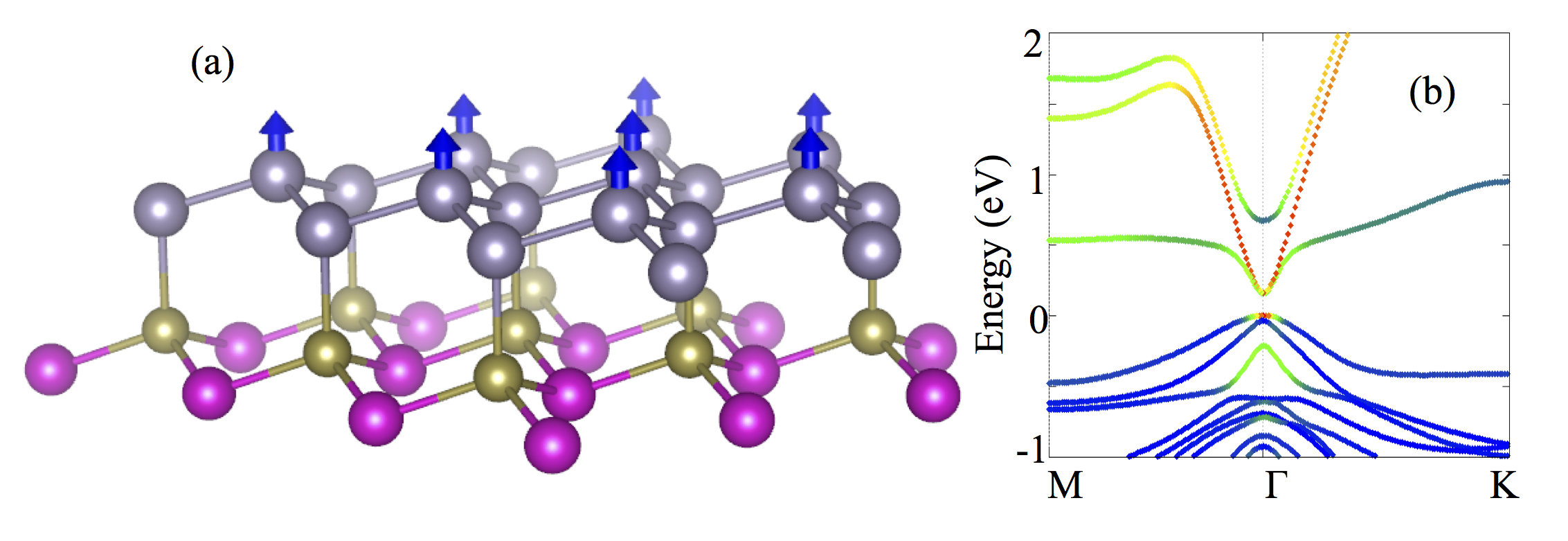}
  \caption{(color online) Stanene grown on the CdTe substrate. (a) Atomic structure. Purple, yellow, and gray spheres represent Cd, Te, and Sn atoms, respectively. Blue arrows illustrate the magnetic moments on Sn atoms of one sublattice. (b) Band structures. The highest valence band at the $\Gamma$ point is dominantly contributed by the Sn-$s$ orbital (highlighted by the red color), which indicates the band inversion in the spin-down channel, topological equivalent to case in Fig.~\ref{fig:band}e.}
\label{fig:CdTe}
\end{figure}

We use the I-passivated stanene as an example to demonstrate the QAH effect in the following discussion. We found a net spin-polarization with the magnetic moment of approximately 1~$\mu_{\rm B}$ per unit cell for the half-I-passivated stanene. The spin charge density is shown {\color{black}in Fig.~\ref{fig:structure}b}. Most of the total magnetic moment is contributed by Sn-$p_z$ states of the unpassivated Sn atom, while the left moment is found to distribute at the other Sn and I sites, which are expected to mediate the exchange coupling between those unpassivated magnetic Sn sites. 
We expect that the FM lattice can be visualized by spin-polarized STM in future experiments.
In the following band structure calculations, we presume the FM order (the stabilization of FM order will be discussed in the following text) as the ground state and investigate topological properties. {\color{black} Before SOC is applied (band structures in Fig.~\ref{fig:band}b and density of states(DOS) in Fig.~\ref{fig:band}c), the $s-p_{xy}$ inversion is removed in the spin-up channel, while it remains in the spin-down channel due to exchange coupling, because Sn-$p_z$ dangling bond states hybridize strongly with Sn-$s$ and Sn-$p_{xy}$ states. The system is half-metallic, for the spin-down channel is gapless owing to the $p_{xy}$ degeneracy at the $\Gamma$ point while the spin-up channel exhibits a gap. When SOC is introduced, such degeneracy is lifted and the system turns into an insulator. Consequently, we realize a single band inversion only in the spin-down channel in the FM insulator, realizing an QAH insulator(Fig.~\ref{fig:band}e). The evolution from QSH to QAH state is illustrated with respect to the exchange field and SOC in Fig.~\ref{fig:band}g. More details on band structures can be found in Ref.\cite{SM} }
At the edge, as a result, the left mover of the QSH state (see Fig.~\ref{fig:edge}a) that corresponds to spin-up channel is suppressed and only the right mover in the spin-down channel remains, which characterizes the QAH effect. Our edge state calculations show a chiral edge state at the boundary of half-passivated stanene, as shown in Fig.~\ref{fig:edge}b. We note that the spin of QAH edge state does not orient totally along the $-z$ direction (out-of-plane), but exhibits a slight $y$ (in-plane) component, because the $xy$-plane of the system does not have a mirror symmetry that can confine the spin along the $z$ axis. It is remarkable that the indirect energy-gap is 0.34~eV, which is far beyond the room temperature. The existence of chiral edge states inside such a large energy gap is a manifestation of the topological property of bulk Bloch states of valence bands. This is characterized by the quantized Hall conductance $\sigma_{xy} \equiv \mathcal{C} e^2 /h$, where $\mathcal{C}$ is an integer known as the Chern number, $h$ is Planck’s constant, and $e$ is the charge of an electron. We obtained $\sigma_{xy} = 3.874045 \times 10^{-5} ~\mathrm{S} = 1 ~e^2 /h$ in the anomalous Hall conductance calculations using the Kubo-Greenwood formula based on the bulk band structure in Fig.~\ref{fig:band}e, and confirmed the topologically nontrivial feature with a non-zero finite Chern number $\mathcal{C}=1$. In addition, we found that the magnetization axis greatly favors the out-of-plane direction by 2~meV for I-passivated stanene, compared to other in-plane directions. Although in-plane magnetization can also induce the QAH effect when it breaks the reflection symmetry~\cite{Liu2013}, the bulk energy-gap is found to be small or even zero (semimetal) in current systems for the in-plane magnetization cases.
The band evolution from QSH to QAH states is further illustrated in Fig.~\ref{fig:band}g, which is calculated based on an effective model in Eq.~\ref{eq:Heff}. Herein, only Sn-$s$ and Sn-$p_{xy}$ bands that are involved in the band inversion are considered for simplicity. We note that the lattice parameters of half-passivated stanene are reduced by 3\% compared to those of fully-passivated stanene. The reduced lattice constants decreases the strength of the $sp$ band inversion from 0.66 to 0.42~eV, which is important to quantitatively understand the band structures. Due to the existence of TR and inversion symmetries, all bands are doubly degenerate in the QSH state. When ferromagnetism exists, these bands exhibit spin splitting due to magnetization. The splitting values are 2$G_s=1.38$~eV and 2$G_p=-0.34$~eV for Sn-$s$ {\color{black}($j=1/2, m_j=\pm 1/2$)} and Sn-$p_{xy}$ {\color{black}($j=3/2, m_j=\pm 3/2$)} bands, respectively. Because the amplitude of 2$G_s$ is much larger than that of 2$G_p$ herein, the $|\mathrm{Sn}-s,\uparrow>$ state is even higher than the $|\mathrm{Sn}-p,\downarrow>$ state inside the conduction bands. The effective Hamiltonian for half-I-passivated stanene can be described by the Bernevig-Hughes-Zhang (BHZ) Hamiltonian~\cite{bernevig2006d} with an additional Zeeman type of coupling~\cite{liu2008},
\begin{eqnarray}
             H &=& H_{\rm {BHZ}}+H_G \label{eq:Heff}\\
 H_{\rm {BHZ}} &=& \left[
                   \begin{array}{cc}
                          h(\bf k) & 0\\
                          0 & h^*(-\bf k)
                   \end{array}
                   \right] \nonumber\\
           H_G &=& \left[
                   \begin{array}{cccc}
                          G_s & 0 & 0 & 0\\
                          0 & G_p & 0 & 0\\
                          0 & 0 & -G_s & 0\\
                          0 & 0 & 0 & -G_p
                   \end{array}
                   \right],\nonumber
\end{eqnarray}
where $h({\bf k})=\epsilon_{\bf k}\mathbb{I}_{2\times2}+M({\bf k})\sigma_z+A(k_x\sigma_x-k_y\sigma_y)$, $M({\bf k})=M_0+B{\bf k}^2$, $\epsilon_{\bf k}=C+D\bf{k}^2$, and $\sigma_{x,y,z}$  are the Pauli matrices. The parameters were obtained by fitting the \textit{ab-initio} band structures in Fig.~\ref{fig:band}e: $M_{0}=-0.2$~eV, $G_{s}=0.69$~eV, $G_{p}=-0.17$~eV, $A=0.6a$~eV$\cdot$\AA, $B=1.5a^{2}$~eV$\cdot$\AA$^2$, $D=1.2a^2$~eV$\cdot$\AA$^2$, $C$ is an arbitrary constant, and $a=4.77$ is the lattice parameter in units of \AA. Due to the large amplitude and opposite signs of these two exchange splitting, i.e. $G_s$ and $G_p$, the band inversion in spin-up channel of $H_{\rm {BHZ}}$ is removed while that in the spin-down channel is enhanced, inducing the QAH effect. 
{\color{black} The opposite signs of $G_s$ and $G_p$ are similar to the mechanism of QAH effect in Mn doped HgTe QWs~\cite{liu2008}. The large amplitude of $G_s$ compared to $G_p$ can be understood from the orbital overlap. The wave function of the $s$-band distributes mainly on the Sn site since it is an anti-bonding state. In contrast, the wave function of the $p_{xy}$-band distributes mainly at the bond center since it is a bonding state. Therefore, $s$-band exhibits much larger overlap with the unpaired $p_z$ state, which is centered at the Sn atom, than the $p_{xy}$-band does.}

\begin{table}[htb]
\centering
\caption{Magnetic properties of stanene and germanene that are half-passivated by Br, Cl, or I atoms. The I passivated stanene and germanene are found to be FM insulators, while other materials prefer  coplanar 120$^{\circ}$ AFM ordering. The effective exchange coupling $J_{eff}$ and mean-field Curie temperature $T^{\rm {MF}}_{\rm C}$ of these FM insulators are estimated from our DFT total energy calculations. When forcing the FM phase, all materials are found to be QAH insulators and corresponding energy gaps ($E_{\rm g}$) are shown in unit of eV.
}
\begin{tabular}{cccccc}
\hline
          &  & Ground state  &  $J_{eff}$(meV)  &  $T^{\rm {MF}}_{\rm C}$(K)& $E_{\rm g}$(eV)\\
\hline
                      & I   & FM   &     7     &      243  &   0.34        \\
Stanene        & Br & AFM &     -     &      -       &    0.36           \\
            	    & Cl &  AFM &    -     &      -       &     0.39          \\
\hline
                     & I  & FM    &     15    &     509 &  0.06          \\
Germanene & Br & AFM &     -       &      -     &  0.21          \\
                    & Cl & AFM &     -       &      -     &   0.20          \\
\hline
\end{tabular}
\label{tab:gap}
\end{table}

As we see from both the charge density and band structure, the unpaired Sn-$p_z$ state strongly couples with other valence and conduction states that are expected to mediate the magnetic coupling. To verify the magnetic ground state, we performed \textit{ab-initio} total energy calculations on FM and antiferromagnetic (AFM) configurations for stanene and germaneness that are passivated with different halides. The magnetization axis of the FM state is set as the $z$ direction, because we found that $z$ direction is more favored than the $xy$ plane in total energy in the FM case. The AFM phase is chosen as the  coplanar noncollinear 120$^{\circ}$ AFM ordering. {\color{black} The ground state is chosen as the one with lowest total energies among FM and AFM.}
We found that I-functionalized stanene and germanene prefer the FM phase, while the other compounds prefer the 120$^{\circ}$ AFM state, as listed in Table~\ref{tab:gap}. Here, the half-I-passivated germanene is also found to be a QAH insulator by the band structure and edge state calculations, similar to the stanene case. Because both FM insulators prefer the $z$ magnetization axis, we can use the effective spin model, $\displaystyle H = - J_{eff}\sum_{<i,j>} S_i S_j$, where $J_{eff}$ is the effective nearest-neighbor exchange coupling, $<i,j>$ indicates that sites $i$ and $j$ are nearest neighbors, and $S_i$ is the spin value at lattice site $i$. By comparing the DFT total energies of the FM phase and a simple strip-like AFM phase, the values of $J_{eff}$ are extracted as 7 (15) meV for stanene (germanene). Next, we can estimate the Curie temperature in a mean field way as $T^{\rm {MF}}_{\rm C}=\frac{2}{3k_{\rm B}} z S(S+1) J_{eff} $, where $\sigma=1/2$, $z=6$ for the triangular lattice, and $k_{\rm B}$ is the Boltzmann constant. It is obtained that $T^{\rm {MF}}_{\rm C} = $ 244 (509) K for stanene (germanene).
Given the empirical relation $\frac{T_{\rm C}}{T^{\rm {MF}}_{\rm {C}}}\approx0.61$ (ref.~\cite{Ashcroft2005}), we estimate that $T_{\rm C}=148 (310)$~K for stanene (germanene), which is still higher than the liquid nitrogen temperature (room temperature). For those materials that favor the 120$^{\circ}$ AFM phase, they turn to be QAH insulators when forcing the out-of-plane FM state, in which the corresponding energy gaps are listed in Table~\ref{tab:gap}. It indicates the possibility that external magnetic field may be applied along the $z$ direction to induce such QAH state. 
{\color{black} We note that the ground states of FM and AFM are related to the subtle balance between competing factors, such as exchange coupling and orbital overlap and so on. The lattice constants of I-passivated stanene and germanene are slightly shorter than those of corresponding Br- and Cl-passivated structures, which is plausibly relevant to the existence of FM order in I-passivated structures.
In addition, the magnetism on current $p$-band based honeycomb lattice may exhibit potential of exotic phenomena such as flat-bands
~\cite{Wu2007,Wu2008,Zhang2014honeycomb}.}

Herein, the ferromagnetism has been realized by controlling the passivation on the Sn and Ge honeycomb lattice. A similar mechanism was introduced to induce room temperature FM order in graphene by passivating one sublattice via hydrogenation.~\cite{JZhou2009} 
 However, such an FM state has yet to be experimentally observed, which is due to the clustering of adatoms and defects on graphene~\cite{Nair2012}.  In order to avoid these technical challenges encountered in graphene, we suggest that the magnetism can also be achieved with the assistance of a proper substrate, instead of halide passivation, for the experimental realization of the QAH effect proposed in stanene and germanene. For example, pristine stanene lattice can be grown on the CdTe or InSb (111) surface owing to the matched lattice and close lattice parameters. As shown in Fig.~\ref{fig:CdTe}a, one Sn sublattice connects to the substrate of CdTe by strong Sn-Te chemical bonds, while the other Sn sublattice buckles in the out-of-plane direction with dangling bonds. Consequently, the unpassivated Sn sublattice exhibits ferromagnetism, which is equivalent to the case of half-I-passivated stanene. It should be noted that the substrate-induced strain slightly affects the band structure of the stanene. We found that a 2\% increase in the CdTe in-plane lattice constant can efficiently produce a topological nontrivial band structure, as shown in Fig.~\ref{fig:CdTe}b. The QAH state is clearly characterized by the $sp$ band inversion in the spin-down channel.

In summary, we propose that the half-passivated stantene and germanene are quantum anomalous Hall systems with a large energy gap. The FM order is realized on the unpassivated sublattice that exhibits dangling bonds with high Curie temperature. Strong coupling between the spin-polarized dangling bond states ($p_z$) and the inherent $s-p_{xy}$  inverted bands opens an considerable energy gap. Inside this bulk gap, gapless chiral edge states emerges to characterize the quantum anomalous Hall effect. Such a half-passivated honeycomb structure can also be realized on an insulating substrate.  {\color{black} 
After the submission of the manuscript, we realize a similar QAH proposal~\cite{Huang2014} on the half-saturated honeycomb lattice wherein model Hamiltonian calculations were performed with only considering $p_z$ orbitals and revealed QAH phase that survives only with $weak$ exchange field.
Their model may be more suitable for graphene and silicene, wherein the $s-p_{xy}$ inversion does not exist. However, for germanene and stanene, the $s-p_{xy}$ inversion becomes an essential feature of the band structure, wherein the multi-orbital feature was appreciated to design QSH and QAH states~\cite{Zhang2014honeycomb}.
Therefore, our calculations that include $s$ and $p_{xyz}$ states can describe stanene and germanene well and demonstrate the QAH state with $strong$ exchange field.}

We thank the helpful discussions with C. Felser, S. Kanugo, C.-X. Liu, Z. Wang, Y. Xu, K. Wu, and Y. Zhou. B.Y. acknowledges financial support from the ERC Advanced Grant (291472) and computing time at HLRN Berlin/Hannover (Germany).

%

\end{document}